\documentclass[sn-mathphys]{sn-jnl}% Math and Physical Sciences Reference Style

\jyear{2021}%

\theoremstyle{thmstyleone}%
\theoremstyle{thmstyletwo}%

\theoremstyle{thmstylethree}%

\begin{document}

\title[Article Title]{Interactive Entanglement in Hybrid Opto-magno-mechanics System}


\author[1]{\sur{Jun Wang}}

\author[1]{\sur{Jing-Yu Pan}}
%% \equalcont{These authors contributed equally to this work.}

\author[1]{\sur{Ya-Bo Zhao}}

\author[1]{\sur{Jun Xiong}}

\author*[1]{\sur{Hai-Bo Wang}}\email{hbwang@bnu.edu.cn}

\affil[1]{\orgdiv{Department of Physics}, \orgname{Beijing Normal University}, \orgaddress{\city{Beijing}, \postcode{100875}, \country{China}}}
%% \street{Street}

\abstract{We present a novel cavity opto-magno-mechanical hybrid system to generate entanglements among multiple quantum carriers, such as magnons, mechanical resonators, and cavity photons in both the optical and microwave domains. Two Yttrium iron garnet (YIG) spheres are embedded in two separate microwave cavities which are joined by a communal mechanical resonator. Because the microwave cavities are separate, the ferromagnetic resonate frequencies of two YIG spheres can be tuned independently, as well as the cavity frequencies. We show that entanglement can be achieved  with experimentally reachable parameters. The entanglement is robust against environmental thermal noise, owing to the mechanical cooling process achieved by the optical cavity. The maximum entanglement among different carriers is achieved by optimizing the parameters of the system. The individual tunability of the separated cavities allows us to independently control the entanglement properties of different subsystems and establish quantum channels with different entanglement properties in one system. This work could provide promising applications in quantum metrology and quantum information tasks. }




\maketitle

\section{Introduction}\label{sec1}

Quantum entanglement is the key resourse for quantum information science, such as quantum computing\cite{QC1,QC2,QC3,QC4}, quantum key distribution\cite{QKD1,QKD2}, quantum secret sharing \cite{QSS1,QSS2,QSS3}, quantum teleportation \cite{QT1}, quantum dense coding\cite{QDC1}, quantum secure direct communication\cite{QSDC1,QSDC2}, and so on. Entanglement has been generated in many systems such as photons\cite{REVphoton}, atoms\cite{REVatom}, and superconductors\cite{REVsuperconduct}. To coherently couple different quantum systems, mechanical oscillators have been widely used \cite{Kippenberg2008, REVoptomechanics}, which lead to the development in the research area known as optomechanics. Ferromagnetic systems, which have been widely studied since 1946\cite{Griffiths1946, Kittel1948, Walker1957}, provide an alternative way to couple different quantum information carriers. Strong coupling between the collective excitation of magnetization in ferrimagnetic materials (which is known as magnon) and photon inside microwave (MW) cavity has been achieved\cite{Soykal2010, Huebl2013}. Yttrium iron garnet (YIG, Y$_3$Fe$_5$O$_{12}$) is an excellent ferrimagnetic material for quantum information processing, owing to its very high spin density and low loss\cite{YIG1,YIG2}. YIG sphere inside the MW cavity provides strong coupling between magnons and cavity photons near the resonance point, which can be flexibly tuned by adjusting the bias magnetic field. Meanwhile, a membranous mechanical resonator can be coupled to a MW cavity as a vibrating capacitor\cite{Vitali2007, Barzanjeh2011, Cai2019}. In addition, direct coupling between magnons and vibration modes of YIG sphere can be achieved by magnetostrictive interaction\cite{Zhang2016}. These magnon-photon-phonon hybrid systems provide a privileged platform for coherently transferring quantum states among different systems. 

The entanglement properties of hybrid opto-magno-mechanical systems have been widely studied at both mean field level\cite{Zhang2016} and full quantum level\cite{Li2018}. Several protocols to generate entanglement among cavity magnomechanics system have been reported \cite{Li2019a, Zhang2019, Li2019, Yu2020, Yu2020l, Yuan2020, Tan2021}. Schemes to entangle two YIG spheres in a single cavity have been proposed in previous work\cite{Li2019}, but to entangle YIG spheres in separate cavities is still a pending problem. The use of separated cavities will make it more convenient to study the frequency-tunable characteristics of entanglement among YIG spheres and MW cavities. 

Distinguished from all previous approaches, we present a hybrid opto-magno-mechanical system that includes two magnons embedded in two separate MW cavities, a mechanical oscillator, and an ancilla optical cavity. Because the microwave cavities are separate, the ferromagnetic resonate frequencies of two YIG spheres can be tuned independently, as well as the cavity frequencies. Entanglements are generated by the nonlinearity of the system such as the magnetic dipole interaction and radiation pressure. Meanwhile, the mechanical oscillator is cooled by the ancilla optical cavity, which plays an important role in obtaining steady state and decresing thermal noise. We consider the quantum fluctuations and solve the system via linearized quantum Langevin equations. We calculate the entanglement properties by solving the Lyapunov equation and calculating the logarithmic negativity\cite{En1,En2,En3}. The results show that our model can yield strong entanglements by optimizing the detunings between driven fields and cavities or magnons. The individual tunability of the separated cavities allows us to independently control the entanglement properties of different subsystems and establish quantum channels with different entanglement properties in one system. Besides, the entanglements are robust against environmental temperature at the millikelvin level.

\section{The Model} \label{basicmodel}
\begin{figure}[!ht]
	\centering
	\includegraphics[width=0.9 \linewidth]{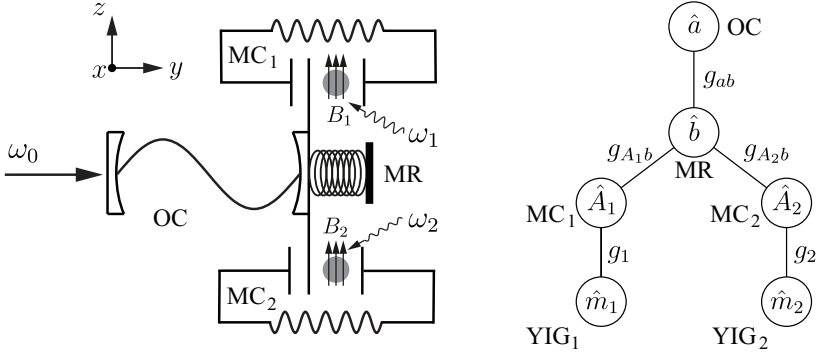}
	\caption{Schematic diagram. A mechanical resonator (MR) is coupled to an optical cavity (OC) and two microwave cavities (MCs). An yttrium iron garnet (YIG) sphere is embedded at the magnetic-field antinode of the microwave cavity. The YIG spheres inside MCs are driven by microwaves $\omega_1$ and $\omega_2$ respectively. At the position of the YIG sphere, the bias magnetic field is along the $z$ direction, the driving magnetic field is along the $y$ direction, and the magnetic field of the cavity mode is along the $x$ direction. }
\end{figure}

The magnon-photon-phonon coupling system is shown in Fig.1. A communal mechanical resonator (MR) is coupled to an optical cavity (OC) and capacitively coupled to two microwave cavities (MC)\cite{Vitali2007, Barzanjeh2011, Cai2019}. An yttrium iron garnet (YIG) sphere is embedded in each microwave cavity. The resonate frequencies of OC and two MCs are $\omega_a$, $\omega_{A1}$, and $\omega_{A2}$, respectively. The ferromagnetic resonate frequencies of two YIG spheres are $\omega_{m_1}$ and $\omega_{m_2}$, which can be tuned by the static bias magnetic field $B_j$ ($j$ = 1, 2) via $\omega_{m_j}=\gamma B_j$ ($\gamma/2\pi$ = 28GHz/T is the gyromagnetic ratio). The MR couples with cavity fields through radiation pressure interaction, with coupling rates $g_{ab}$ (MR-OC), $g_{A1b}$ (MR-MC1), and $g_{A2b}$ (MR-MC2). The magnons inside MCs couple with cavity fields through magnetic dipole interaction, with coupling rates $g_{1}$ and $g_{2}$. The OC is driven by an optical field $\omega_0$, while the YIG spheres inside MCs are driven by microwaves $\omega_1$ and $\omega_2$  respectively. The direct coupling between the YIG sphere and the microwave driving field is adopted in previous work \cite{You2016, You2018, Li2018}. The decay rates of OC, two MCs, and two magnons are $\kappa_a$, $\kappa_{A_i}$, and $\kappa_{m_i}$ ($i$=1,2). The total Hamiltonian is:
\begin{align}
	H=&\hbar\omega_a a^{\dag}a +\hbar\sum_{i=1,2}(\omega_{m_i}m_i^{\dag}m_i+\omega_{A_i}A_i^{\dag}A_i)+\hbar\omega_b b^{\dag}b -g_{ab}a^{\dag}a(b+b^{\dag}) \nonumber\\
	&+\hbar\sum_{i=1,2}[g_i(A_i+A_i^{\dag})(m_i+m_i^{\dag})-g_{Aib}A_i^{\dag}A_i(b+b^{\dag})] \nonumber\\
	&+i\hbar\Omega_0(a^{\dag}e^{-i\omega_0t}-ae^{i\omega_0t})+i\hbar\sum_{i=1,2}\Omega_i(m_i^{\dag}e^{-i\omega_it}-m_ie^{i\omega_it}),
\end{align}
where $a$($a^{\dag}$), $b$($b^{\dag}$), $A_i$($A_i^{\dag}$), and $m_i$($m_i^{\dag}$) are the creation (annihilation) operators for the optical cavity mode, the mechanical mode, the $i$-th microwave cavity mode, and the $i$-th magnon mode, respectively. The Rabi frequencies $\Omega_0$, $\Omega_1$, and $\Omega_2$ denote the strength of the driven fields $\omega_0$, $\omega_1$, and $\omega_2$, respectively. 

In the rotating frame with respect to $\omega_0a^{\dag}a+\omega_1m_1^{\dag}m_1+\omega_2m_2^{\dag}m_2+\omega_1A_1^{\dag}A_1+\omega_2A_2^{\dag}A_2$ and applying the rotating-wave approximation $(A_i+A_i^{\dag})(m_i+m_i^{\dag})\approx A_i m_i^{\dag}+A_i^{\dag}m_i$ (when $\omega_{A_i}, \omega_{mi}\gg g_i, \kappa_{A_i}, \kappa_{m_i}$), the effective Hamiltonian is: 
\begin{align}
	H=&\hbar\Delta_{a_0} a^{\dag}a +\hbar\sum_{i=1,2}(\Delta_{m_i}m_i^{\dag}m_i+\Delta_{A_i}A_i^{\dag}A_i)+\hbar\omega_b b^{\dag}b -g_{ab}a^{\dag}a(b+b^{\dag}) \nonumber\\
	&+\hbar\sum_{i=1,2}[g_i(A_im_i^{\dag}+A_i^{\dag}m_i)-g_{Aib}A_i^{\dag}A_i(b+b^{\dag})]+i\hbar\Omega_0(a^{\dag}-a) \nonumber\\
	&+i\hbar\sum_{i=1,2}\Omega_i(m_i^{\dag}-m_i),
\end{align}
where $\Delta_{a_0}=\omega_a-\omega_0$, $\Delta_{mi}=\omega_{m_i}-\omega_i$, and $\Delta_{Ai}=\omega_{A_i}-\omega_i$ denote the detunings of the driven fields. The quantum Langevin equations (QLEs) of the system are:
	\begin{align}
	\dot{a}=&-(i\Delta_{a_0}+\kappa_a)a+ig_{ab}(b+b^{\dag})a+\Omega_0+\sqrt{2\kappa_a}a^{in}, \\
	\dot{b}=&-(i\omega_b+\kappa_b)b+ig_{ab}a^{\dag}a+ig_{A_1b}A_1^{\dag}A_1+ig_{A_2b}A_2^{\dag}A_2+\sqrt{2\kappa_b}b^{in}, \\
	\dot{m_i}=&-(i\Delta_{m_i}+\kappa_{m_i})m_i-ig_iA_i+\Omega_i+\sqrt{2\kappa_{m_i}}m_i^{in}, \\
	\dot{A_i}=&-(i\Delta_{A_i}+\kappa_{A_i})A_i+ig_{A_ib}(b+b^{\dag})A_i-ig_im_i+\sqrt{2\kappa_{A_i}}A_i^{in},
\end{align}
where $a^{in}$, $b^{in}$, $m_i^{in}$, and $A_i^{in}$ are input noise operators for the optical cavity mode, the mechanical mode, the magnon modes, and the microwave cavity modes, respectively, which are characterized by the following correlation functions:
\begin{align}
	\langle a^{in}(t)a^{in\dag}(t')\rangle=&[N_a(\omega_a)+1]\delta(t-t'), \\
	\langle b^{in}(t)b^{in\dag}(t')\rangle=&[N_b(\omega_b)+1]\delta(t-t'), \\
	\langle m_i^{in}(t)m_i^{in\dag}(t')\rangle=&[N_{m_i}(\omega_{m_i})+1]\delta(t-t'), \\
	\langle A_i^{in}(t)a^{in\dag}(t')\rangle=& [N_{A_i}(\omega_{A_i})+1]\delta(t-t'),
\end{align}
where $N_i(\omega_i)=\{\exp [(\hbar\omega_i/k_B T)-1]\}^{-1}$ ($i$= $a$,  $A_1$,$A_2$,$m_1$,$m_2$, $b$) are the Bose-Einstein distribution of thermal photons, magnons and phonons. 

The QLEs can be linearized in the strongly driven approximation, namely, for $a$, $b$, $m_i$, and $A_i$, their steady state amplitude $\langle O\rangle$ ($O=a,b,m_1,m_2,A_1,A_2$) is much larger than their fluctuation $\delta O$. Substitute $O=\langle O\rangle+\delta O$ into the QLEs and ignore the higher-order terms of $\delta O$, we got the steady state solutions:
\begin{align}
	\langle a\rangle=&\frac{\Omega_0}{i\tilde{\Delta}_{a_0}+\kappa_a} \label{eq11}, \\
	\langle m_i\rangle=&\frac{(i\tilde{\Delta}_{A_i}+\kappa_{A_i})\Omega_i}{g_i^2+(i\Delta_{m_i}+\kappa_{m_i})(i\tilde{\Delta}_{A_i}+\kappa_{A_i})} \label{eq12}, \\
	\langle A_i\rangle=&\frac{g_i\langle m_i\rangle}{-\tilde{\Delta}_{A_i}+i\kappa_{A_i}} \label{eq13}, \\
	\langle b\rangle=&\frac{g_{ab}\lvert\langle a\rangle\rvert^2+g_{A_1b}\lvert\langle A_1\rangle\rvert^2+g_{A_2b}\lvert\langle A_2\rangle\rvert^2}{\omega_b-i\kappa_b} \label{eq14}, 
\end{align}
where $\tilde{\Delta}_{a_0}=\Delta_{a_0}-g_{ab}(\langle b \rangle+\langle b^{\dag} \rangle)$ and $\tilde{\Delta}_{A_i}=\Delta_{A_i}-g_{A_ib}(\langle b \rangle+\langle b^{\dag} \rangle)$ are the effective detunings of the optical cavity and two microwave cavities. We also get the equations of the quadrature fluctuations $\delta X_{O}$ and $\delta Y_{O}$ ($\delta X_{O}=(\delta O+\delta O^{\dag})/\sqrt{2}$ and $\delta Y_{O}=(\delta O-\delta O^{\dag})/i\sqrt{2}$, with $O=a,b,m_1,m_2,A_1,A_2$):
\begin{align}
	\dot{u}(t)=A{u}(t)+n(t),
\end{align}
where $u(t)=[\delta X_{a}(t)$, $\delta Y_{a}(t)$, $\delta X_{b}(t)$, $\delta Y_{b}(t)$, $\delta X_{A_1}(t)$, $\delta Y_{A_1}(t)$, $\delta X_{m_1}(t)$, $\delta Y_{m_1}(t)$, $\delta X_{A_2}(t)$, $\delta Y_{A_2}(t)$, $\delta X_{m_2}(t)$, $\delta Y_{m_2}(t)]^T$, $n(t)=[\sqrt{2\kappa_a}X_{a}^{in}(t)$, $\sqrt{2\kappa_a}Y_{a}^{in}(t)$, $\sqrt{2\kappa_b}X_{b}^{in}(t)$, $\sqrt{2\kappa_b}Y_{b}^{in}(t)$, $\sqrt{2\kappa_{A_1}}X_{A_1}^{in}(t)$, $\sqrt{2\kappa_{A_1}}Y_{A_1}^{in}(t)$, $\sqrt{2\kappa_{m_1}}X_{m_1}^{in}(t)$, $\sqrt{2\kappa_{m_1}}Y_{m_1}^{in}(t)$, $\sqrt{2\kappa_{A_2}}X_{A_2}^{in}(t)$, $\sqrt{2\kappa_{A_2}}Y_{A_2}^{in}(t)$, $\sqrt{2\kappa_{m_2}}X_{m_2}^{in}(t)$, $\sqrt{2\kappa_{m_2}}Y_{m_2}^{in}(t)]^T$, and
\addtocounter{MaxMatrixCols}{10}
{\footnotesize
	\begin{align}\label{eq16}
		A=\begin{pmatrix}
			-\kappa_a & \tilde{\Delta}_{a_0} & -2G^{I}_{ab} & 0 & 0 & 0 & 0 & 0 & 0 & 0 & 0 & 0 \\
			-\tilde{\Delta}_{a_0} & -\kappa_a & 2G^{R}_{ab} & 0 & 0 & 0 & 0 & 0 & 0 & 0 & 0 & 0 \\
			0 & 0 & -\kappa_b & \omega_b & 0 & 0 & 0 & 0 & 0 & 0 & 0 & 0 \\
			2G^{R}_{ab} & 2G^{I}_{ab} & -\omega_b & -\kappa_b & 2G^R_{A_1b} & 2G^I_{A_1b} & 0 & 0 & 2G^R_{A_2b} & 2G^I_{A_2b} & 0 & 0 \\
			0 & 0 & -2G^I_{A_1b} & 0 & -\kappa_{A_1} & \tilde{\Delta}_{A_1} & 0 & g_1 & 0 & 0 & 0 & 0 \\
			0 & 0 & 2G^R_{A_1b} & 0 & -\tilde{\Delta}_{A_1} & -\kappa_{A_1} & -g_1 & 0 & 0 & 0 & 0 & 0 \\
			0 & 0 & 0 & 0 & 0 & g_1 & -\kappa_{m_1} & \Delta_{m_1} & 0 & 0 &0  & 0 \\
			0 & 0 & 0 & 0 & -g_1 & 0 & -\Delta_{m_1} & -\kappa_{m_1} & 0 & 0 & 0 & 0 \\
			0 & 0 & -2G^I_{A_2b} & 0 & 0 & 0 & 0 & 0 & -\kappa_{A_2} & \tilde{\Delta}_{A_2} & 0 & g_2 \\
			0 & 0 & 2G^R_{A_2b} & 0 & 0 & 0 & 0 & 0 & -\tilde{\Delta}_{A_2} & -\kappa_{A_2} & -g_2 & 0 \\
			0 & 0 & 0 & 0 & 0 & 0 & 0 & 0 & 0 & g_2 & -\kappa_{m_2} & \Delta_{m_2} \\
			0 & 0 & 0 & 0 & 0 & 0 & 0 & 0 & -g_2 & 0 & -\Delta_{m_2} & -\kappa_{m_2}
		\end{pmatrix}
	\end{align}
}

Here $G^R_{ab}=g_{ab}Re\langle a \rangle$, $G^I_{ab}=g_{ab}Im\langle a \rangle$, $G^R_{A_ib}=g_{A_ib}Re\langle A_i \rangle$, and $G^I_{A_ib}=g_{A_ib}Im\langle A_i \rangle$are the effective coupling rates.

Owing to the linearized dynamics of the QLEs, the Gaussian nature of the quantum noises will be preserved. Thus the steady state of the
quantum fluctuations is a continuous variable (CV) six-mode Gaussian state characterized by a $12\times12$ covariance matrix (CM) $V$: $V_{ij}(t,t')=\frac{1}{2}\langle u_i(t)u_j(t')+u_j(t')u_i(t)\rangle$, which can be obtained by solving the Lyapunov equation \cite{lyap1, lyap2}: 
\begin{align}
	AV+VA^T=-D, 
\end{align}
where the diffusion matrix $D$=diag$[\kappa_a(2N_a+1)$, $\kappa_a(2N_a+1)$, $\kappa_b(2N_b+1)$, $\kappa_b(2N_b+1)$, $\kappa_{A_1}(2N_{A_1}+1)$, $\kappa_{A_1}(2N_{A_1}+1)$, $\kappa_{m_1}(2N_{m_1}+1)$, $\kappa_{m_1}(2N_{m_1}+1)$, $\kappa_{A_2}(2N_{A_2}+1)$, $\kappa_{A_2}(2N_{A_2}+1)$, $\kappa_{m_2}(2N_{m_2}+1)$, $\kappa_{m_2}(2N_{m_2}+1)]$ is defined through $D_{ij}\delta(t-t')=\frac{1}{2}\langle N_i(t)N_j(t')+N_j(t')N_i(t)\rangle$

The bipartite entanglement of subsystems $s_1$ and $s_2$ can be measured by the logarithmic negativity $E_N$\cite{criteria2007}:
\begin{align}
	E_N=\max[0,-\ln 2\tilde{\nu}_{-}], 
\end{align}
where $\tilde{\nu}_{-}=\min \mathrm{eig}\lvert i\Omega PC_{s_1s_2}P\rvert$ ($\Omega=\bigoplus_{j=1}^2 i\sigma_y$) is the symplectic matrix, $\sigma_y$ is the y-Pauli matrix, $P=diag(1,-1,1,1)$, and $C_{s_1s_2}$ is a part of the CM matrix that describes subsystems $s_1$ and $s_2$.

\section{Entanglement Analysis} \label{sec3}

\begin{figure}[!ht]
	\centering
	\includegraphics[width=0.49\linewidth]{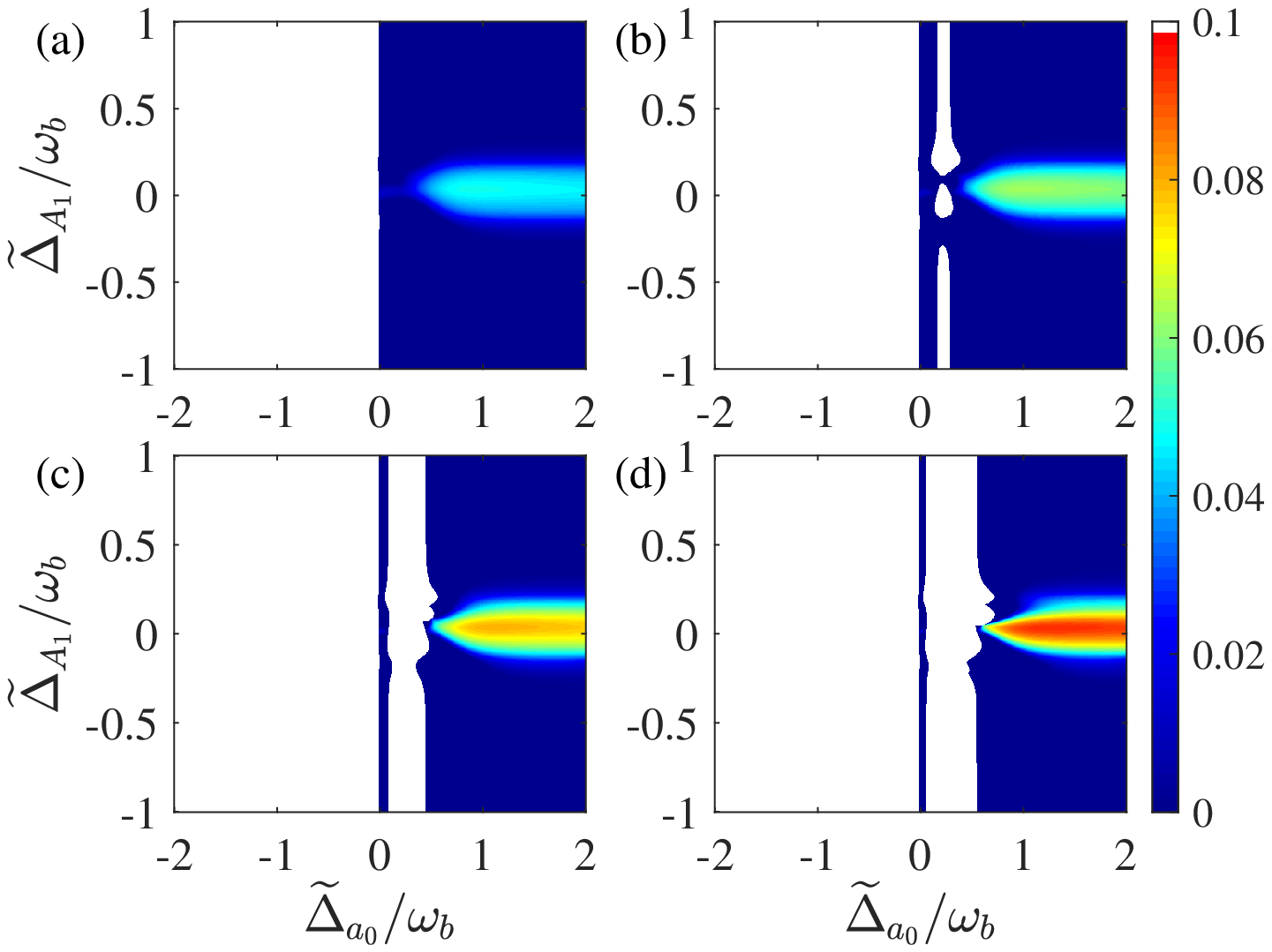}
	\includegraphics[width=0.49\linewidth]{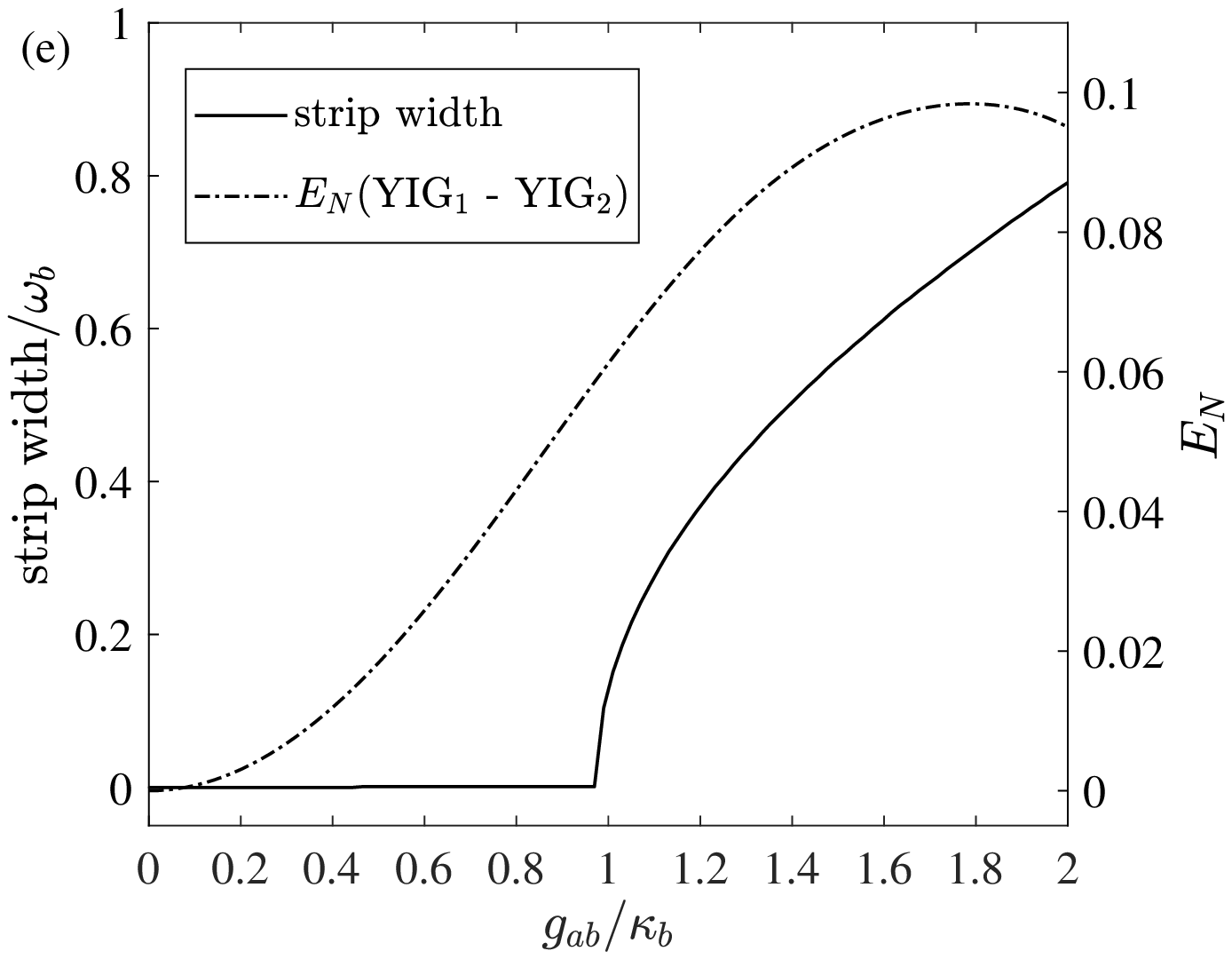}
	\caption{(a)-(d) Steady condition of the system. The system is unsteady in the white area, while the rest part of the figure shows the entanglement $E_N$ between two magnons $m_1$ and $m_2$ versus the effective detunings $\tilde{\Delta}_{a_0}$ and $\tilde{\Delta}_{A_1}$ (with $\tilde{\Delta}_{A_1}$=$\tilde{\Delta}_{A_2}$= $\Delta_{m_1}$=$\Delta_{m_2}$). Corresponding optomechanical coupling rates of the optical cavity $g_{ab}$ are: (a) 0.8 $\kappa_b$, (b)  $\kappa_b$, (c) 1.2 $\kappa_b$, (d) 1.4 $\kappa_b$. (e) The width of the vertical unsteady strip in (a)-(d) and the entanglement at the point $(\tilde{\Delta}_{a_0},\tilde{\Delta}_{A_1})$ = $(\omega_b, 0)$ versus $g_{ab}$. } \label{fig2}
\end{figure}

\begin{figure}[!ht]
	\centering
	\includegraphics[width=0.6\linewidth]{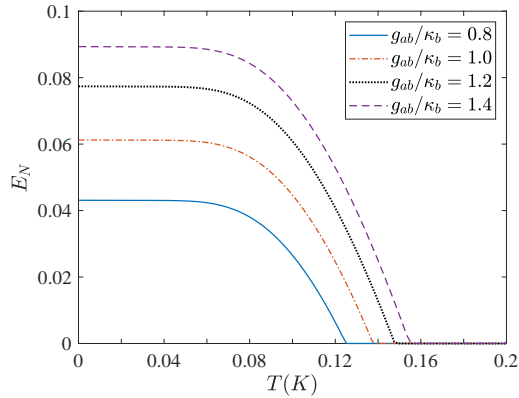}
	\caption{The robustness of entanglement against environmental temperature. The four lines in the figure are the entanglement $E_N$ between magnons of the two YIG spheres with different optomechanical coupling rate of the optical cavity $g_{ab}$. The detuning $\tilde{\Delta}_{a_0}=\omega_b$ and other detunings are zero. } \label{fig3}
\end{figure}

In this section, we analyze the entanglement between different components of the system in the steady state. For the complex hybrid coupling system including three driving sources, it's important to analyze the steady-state condition. The criterion of stability is that all of the eigenvalues (real parts) of the drift matrix A in Eq. (\ref{eq16}) are negative, as a result of the Routh–Hurwitz criterion\cite{Stable2}. In Fig.\ref{fig2} we analyze the steady-state condition of the system with the following experimentally feasible parameters : $\omega_a/2\pi$ = 370 THz, ($\omega_{A_1}$, $\omega_{A_2}$, $\omega_{m_1}$, $\omega_{m_2}$)$/2\pi$ = (10, 10, 10, 10) GHz, $\omega_{b}/2\pi$ = 10MHz, ($\kappa_{a}$, $\kappa_{A_1}$, $\kappa_{A_2}$, $\kappa_{m_1}$, $\kappa_{m_2}$) = (0.4, 0.1, 0.1, 0.1, 0.1)$\omega_{b}$, $\kappa_{b}/2\pi$ = 100Hz, and $g_{1}/2\pi$ = $g_{2}/2\pi$ = 1.7MHz \cite{Zhang2016,Li2018,Yu2020l}, $\Omega_0$ = $1.43\times 10^{12}$ Hz, and $\Omega_1=\Omega_2=7.13\times 10^{14}$ Hz \cite{Li2018}. The result indicates that the detunings of the optical cavity and MW cavity play different roles in the steady condition. This is because the optical light frequency leads to a higher optomechanical coupling rate than the microwave cavity\cite{REVoptomechanics}: 
\begin{align}
	g_0=Gx_{\mathrm{ZPF}}=\omega_{\mathrm{cav}}x_{\mathrm{ZPF}}/L, 
\end{align}
where $g_0$ is the vacuum optomechanical coupling rate, $\omega_{\mathrm{cav}}$ is the cavity frequency, $L$ is the cavity length, and $x_{\mathrm{ZPF}}$ is the zero-point fluctuation amplitude of the mechanical oscillator. Fig.\ref{fig2} shows that the stability of the system is mainly determined by the optical cavity. The red detuned optical cavity ($\tilde{\Delta}_{a_0}>0$) leads to the cooling process of the mechanical resonator and increases the robustness against temperature accordingly. An unsteady white strip appears in the red detuned area when $g_{ab}$ is larger than the dissipation rate of the mechanical oscillator $\kappa_b$. As $g_{ab}$ increases, the unsteady strip becomes wider. This is because the strong optomechanical coupling rate compared with the dissipation rate $\kappa_b$ of the mechanical oscillator will accumulate the energy of the driving field and bring the system to an unsteady state. Fig.\ref{fig2} also shows that the entanglement increases with $g_{ab}$ because the larger optomechanical coupling rate of the optical cavity leads to a stronger cooling process. Hereafter this text we take the value of $g_{ab}$ as 1.2$\kappa_b$ in order to avoid instability in the parameter regime that we investigate. According to the distribution $N_i(\omega_i)=\{\exp [(\hbar\omega_i/k_B T)-1]\}^{-1}$, the mechanical oscillator has larger thermal noise than cavities and magnons owing to its lower eigen energy $\hbar\omega_b$. Fig.\ref{fig3} shows the robustness against environmental temperature. The entanglement remains constant until 60 mK and survives up to 120 mK. Fig.\ref{fig3} also indicates that the larger optomechanical coupling rate of the optical cavity $g_{ab}$ increases the robustness against temperature because it enhances the cooling process and decreases the thermal noise. 

\begin{figure}[!ht]
	\centering
	\includegraphics[width=0.6\linewidth]{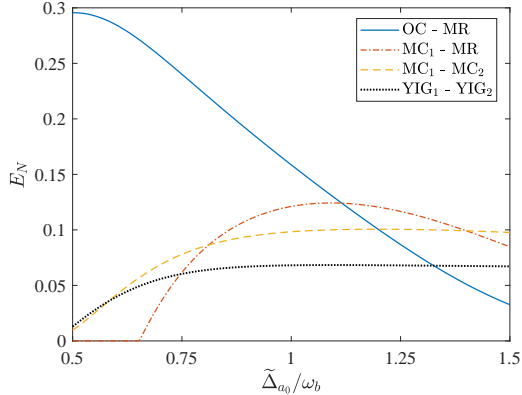} 
	\caption{Entanglement $E_N$ versus the detuning of optical driving field $\tilde{\Delta}_{a_0}$ while $\tilde{\Delta}_{A_1}$=$\tilde{\Delta}_{A_2}$=$\Delta_{m_1}$=$\Delta_{m_2}$=0. Solid line: $E_N$ between optical cavity field and MR; dash-solid line: $E_N$ between MW cavity (MC$_1$) field and MR; dash line: $E_N$ between two MW cavity fields; dot line: $E_N$ between magnons of two YIG spheres.} \label{fig4}
\end{figure}

\begin{figure}[!ht]
	\centering
	\includegraphics[width=0.7\linewidth]{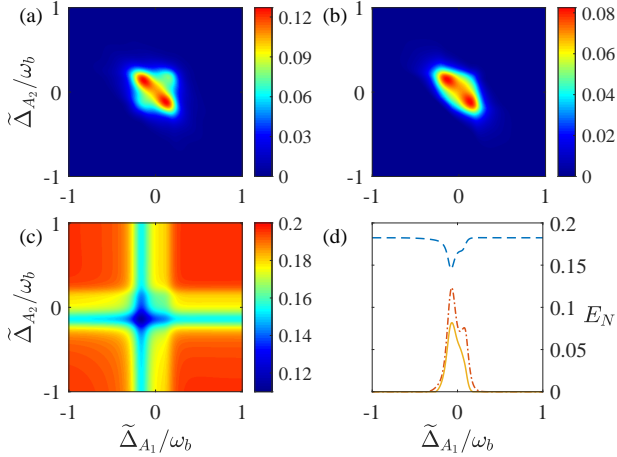}
	\caption{(a)-(c): Entanglement $E_N$ versus $\tilde{\Delta}_{A_1}$ and $\tilde{\Delta}_{A_2}$ while $\tilde{\Delta}_{a_0}$=$\omega_b$, $\tilde{\Delta}_{A_1}$=$\Delta_{m_1}$, and $\tilde{\Delta}_{A_2}$=$\Delta_{m_2}$. (a) $E_N$ between cavity modes of MC$_1$ and MC$_2$; (b) $E_N$ between magnons of YIG$_1$ and YIG$_2$; (c) $E_N$ between OC and MR. (d) Entanglement $E_N$ versus the detuning of MW driving field $\tilde{\Delta}_{A1}$ while $\tilde{\Delta}_{a_0}$=$\omega_b$, $\tilde{\Delta}_{A_1}$=$\Delta_{m_1}$, and $\tilde{\Delta}_{A_2}$=$\Delta_{m_2}$=0.15$\omega_b$. Dash-solid line: $E_N$(MC$_1$-MC$_2$); solid line: $E_N$(YIG$_1$-YIG$_2$); dash line: $E_N$(OC-MR). } \label{fig5}. 
\end{figure}

The components of the hybrid coupling system are individually tunable. Therefore it's important to analyze the effect of detunings on the entanglement of the system. Fig.\ref{fig4} indicates that the optical cavity detuned at the red sideband $\tilde{\Delta}_{a_0}=\omega_b$ materializes the best cooling process and achieves the maximum MR - MC, MC - MC, and magnon - magnon entanglements, which have been discussed earlier. Fig.\ref{fig5}(d) shows the variation of three types of bipartite entanglements when the frequencies of the MW-driven fields change. The complementary variation of $E_N$(OC-MR) and $E_N$(MC$_1$-MC$_2$) and $E_N$(YIG$_1$-YIG$_2$) indicates that entanglement is transferred from the optomechanical subsystem in the optical domain to the opto-magno-mechanical subsystem in the MW domain. This phenomenon is pronounced when the MW cavities are resonantly driven because a resonantly driven cavity has the maximum average cavity photon number $n_{\mathrm{cav}}$ and the maximum effective optomechanical coupling strength $g=g_0 \sqrt{n_{\mathrm{cav}}}$. The complementary variation of entanglements is demonstrated in more detail in Fig.\ref{fig5}(a)-(c), where we assume that the MW cavity and the inside YIG sphere have the same eigenfrequency. The case that the MW cavity and the inside YIG sphere are tuned independently is discussed next.

\begin{figure}[!ht]
	\centering
	\includegraphics[width=0.8\linewidth]{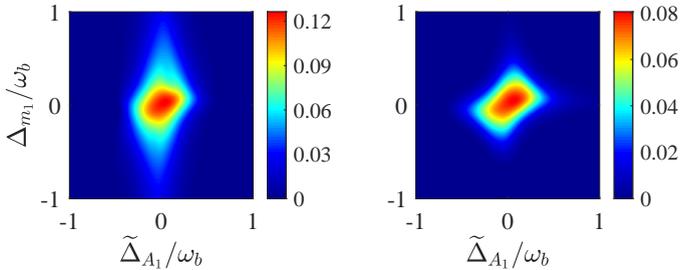}
	\caption{Entanglement $E_N$ versus the detuning of the MW cavity $\tilde{\Delta}_{A_1}$ and the detuning of the inside YIG sphere $\Delta_{m_1}$ while $\tilde{\Delta}_{A_2}$ and $\Delta_{m_2}$ remain zero. (a) $E_N$ between cavity modes of MC$_1$ and MC$_2$; (b) $E_N$ between magnons of YIG$_1$ and YIG$_2$. } \label{fig6}
\end{figure}

The ferromagnetic resonant frequency of YIG spheres determined by the bias magnetic field can be tuned independently of the cavity frequency. Thus for a certain MW driving frequency, the detuning of the MW cavity $\tilde{\Delta}_{A_1}$ and the detuning of the inside YIG sphere $\Delta_{m_1}$ can be different. The variation of entanglements $E_N$ versus $\tilde{\Delta}_{A_1}$ and $\Delta_{m_1}$ is demonstrated in Fig.\ref{fig6}, where the detunings of the other MW cavity $\tilde{\Delta}_{A_2}$ and $\Delta_{m_2}$ remain zero. Fig.\ref{fig6}(a) and (b) show that the maximum cavity-cavity and magnon-magnon entanglements are achieved when the driving MW field resonates with the MW cavity and the internal magnon simultaneously.

\begin{figure}[!ht]
	\centering
	\includegraphics[width=1\linewidth]{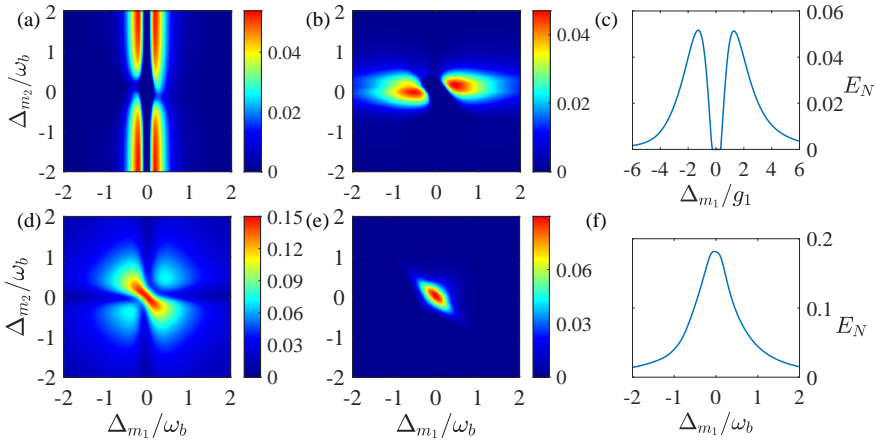}
	\caption{Entanglement $E_N$ versus the detuning of the two YIG spheres $\Delta_{m_1}$ and $\Delta_{m_2}$. (a) $E_N$ between MC$_1$ and YIG$_1$; (b) $E_N$ between MC$_1$ and YIG$_2$; (c) The double-peak structure of $E_N$(MC$_1$-YIG$_1$) while $\Delta_{m_2}$=$\omega_b$; (d)$E_N$ between MW cavity MC$_1$ and MC$_2$; (e) $E_N$ between YIG$_1$ and YIG$_2$; (f) $E_N$ between MC$_1$ and MR while $\Delta_{m_2}$=$\omega_b$. } \label{fig7}
\end{figure}

Our scheme, distinguished from previous works, inserts two YIG spheres into two separate MW cavities respectively. Therefore the ferromagnetic resonate frequencies determined by the static bias magnetic fields can be tuned individually, as well as the cavity frequencies. In Fig.\ref{fig7} we present the variation of entanglements versus different resonate frequencies of YIG spheres while the cavity eigenfrequencies remain constant. Fig.\ref{fig7}(a) shows the entanglement between the cavity field of MC${_1}$ and the magnon of YIG${_1}$. In Fig.\ref{fig7}(a), along the $x$ direction there is a double-peak structure of the entanglement $E_N$(MC$_1$-YIG$_1$). This structure is presented in detail in Fig.\ref{fig7}(c), where the peaks exist at $\Delta_{m_1}\approx\pm 1.24g_1$. The double-peak structure can be explained by the Rabi split. For the simple coupling model with Hamiltonian $H/\hbar=\omega_{A_1}A_1^{\dag}A_1+\omega_{m_1}m_1^{\dag}m_1+g_1(A_1^{\dag}m_1+A_1m_1^{\dag})$ and taking $\omega_{A_1}=\omega_{m_1}$ for the sake of simplicity, the Hamiltonian can be diagonalized by the supermode operators $c_{\pm}=(A_1\pm m_1)$: $H/\hbar=\omega_{+}c_{+}^{\dag}c_{+}+\omega_{-}c_{-}^{\dag}c_{-}$, with supermode eigenfrequencies $\omega_{\pm}=\omega_{A_1}\pm g_1$. The coupling system is resonantly driven when the driving field frequency $\omega_1=\omega_{\pm}$, thus the maximal local cavity-magnon entanglement is achieved at $\Delta_{m_1}\approx\pm g_1$. The deviation from $\pm g_1$ is because of the additional coupling with MR and MC$_2$. In addition, along the $y$ direction in Fig.\ref{fig7}(a) there is a dip around $\Delta_{m_2}=0$, while in Fig.\ref{fig7}(b) the maximal cross-cavity entanglement $E_N$(MC$_1$-YIG$_2$) is achieved around $\Delta_{m_2}=0$. The complementary variation of $E_N$(MC$_1$-YIG$_1$) and $E_N$(MC$_1$-YIG$_2$) indicates that the entanglement in the local cavity is transferred to the cross-cavity subsystem. Fig.\ref{fig7}(d) and (e) present that the maximal entanglements $E_N$(MC$_1$-MC$_2$) and $E_N$(YIG$_1$-YIG$_2$) are achieved around $\Delta_{m_1}=\Delta_{m_2}=0$, because MC$_1$ and MC$_2$ are connected by MR and the maximal entanglement $E_N$(MC$_1$-MR) is achieved when $\Delta_{m_1}=0$ (as shown in Fig.\ref{fig7}(f)). The four types of bipartite entenglement in Fig.\ref{fig7} have different features when independently tuning $\Delta_{m_1}$ and $\Delta_{m_2}$, which indicates that by independently tuning the two YIG spheres in separated MW cavities one can establish quantum channels with different entanglement properties in one system.

\begin{figure}[!ht]
	\centering
	\includegraphics[width=0.49\linewidth]{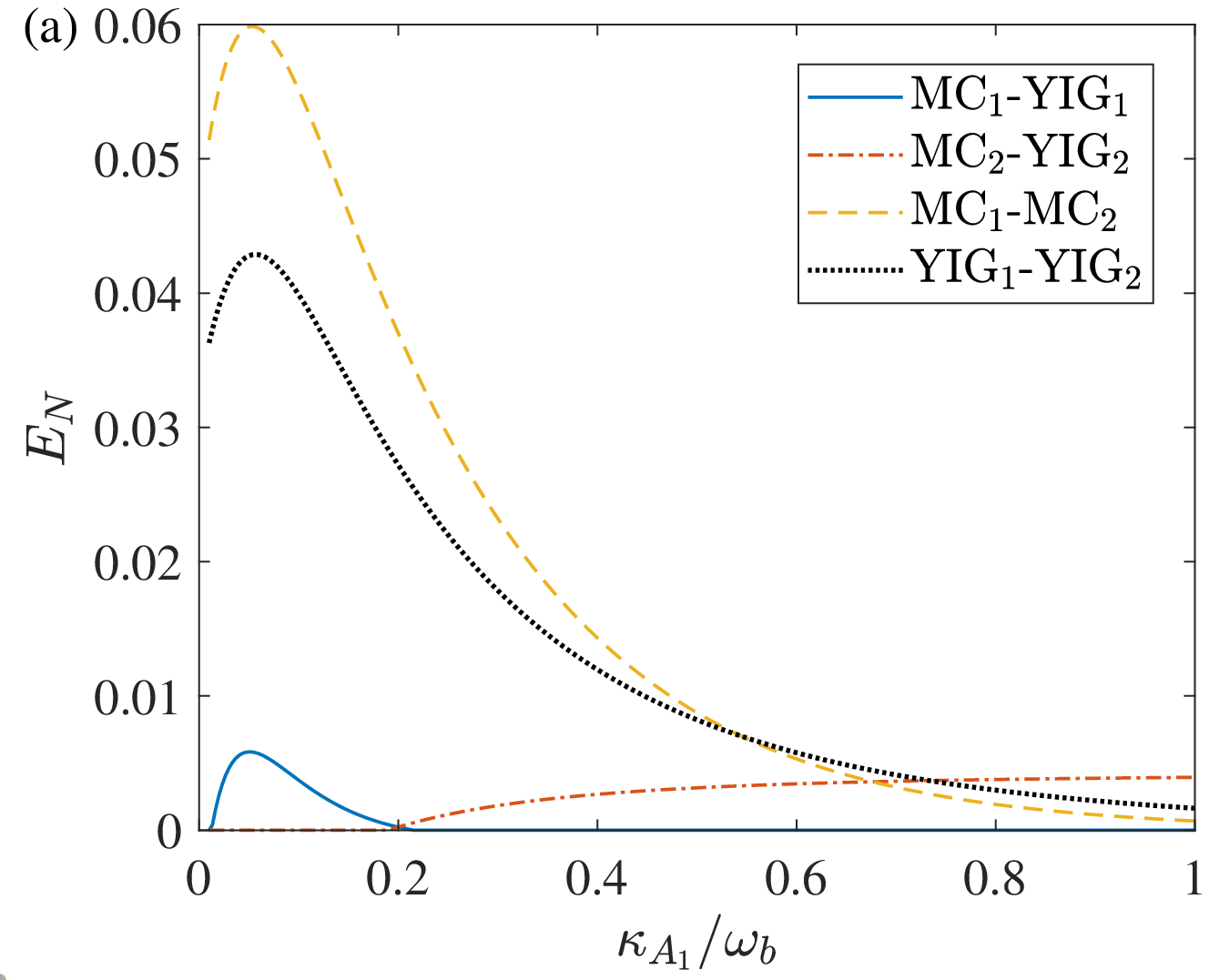}
	\includegraphics[width=0.49\linewidth]{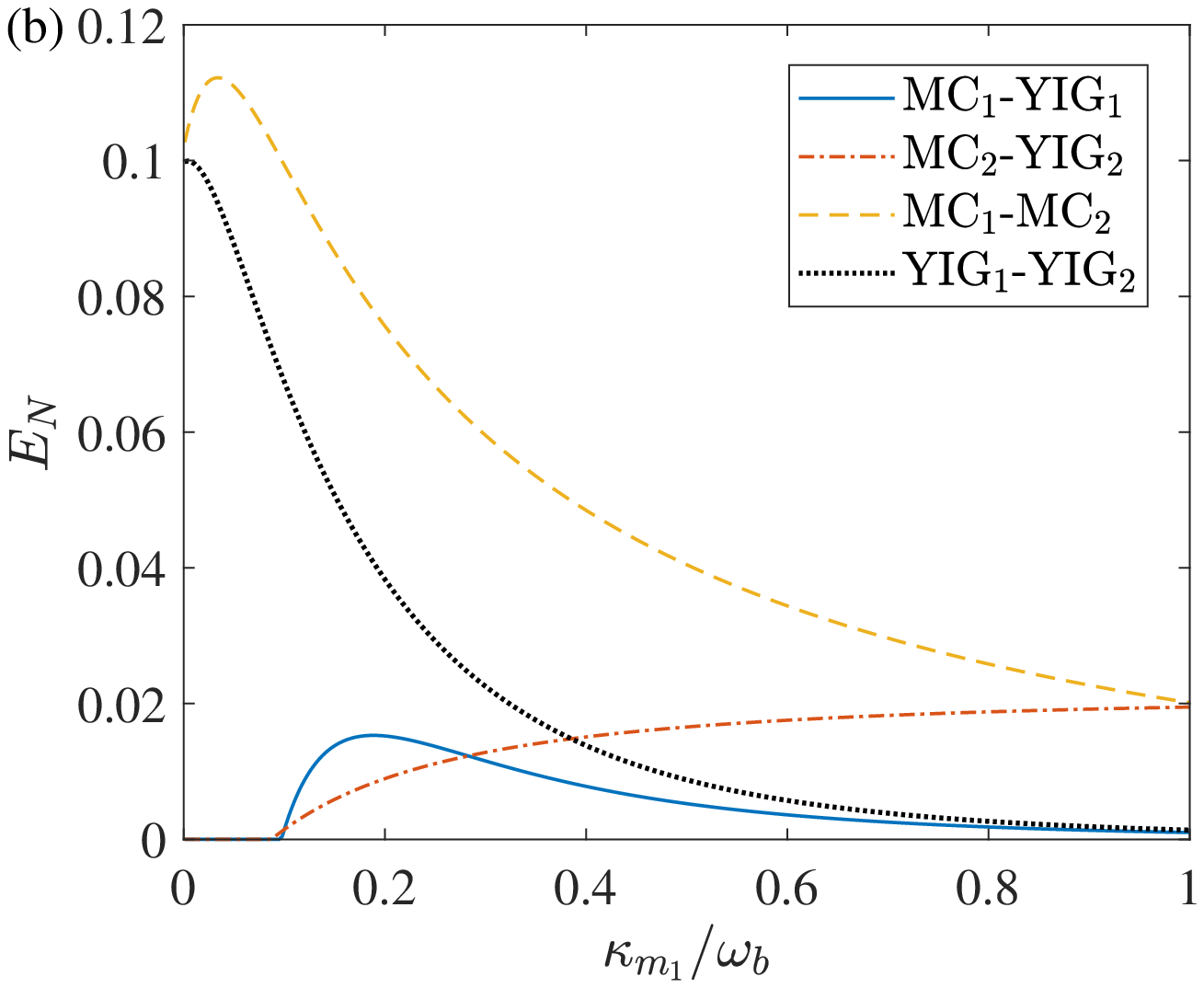}
	\caption{(a) Entanglement $E_N$ versus the dissipation rate $\kappa_{A_1}$ of MC$_1$ while $\kappa_{A_2}$=0.2$\omega_b$. (b) Entanglement $E_N$ versus the dissipation rate $\kappa_{m_1}$ of YIG$_1$ sphere while $\kappa_{m_2}$=0.1$\omega_b$. } \label{fig8}
\end{figure}

The dissipation rate of cavity is determined by the quality factor $Q=\omega_{\mathrm{cav}}/\kappa$ while the dissipation rate of magnon is determined by the shape of YIG and the Gilbert damping process \cite{rameshti2022cavity}. In Fig.\ref{fig8}(a) we analyze the effect of the cavity dissipation rate $\kappa_{A_1}$ while $\kappa_{A_2}$ remains 0.2$\omega_b$. The entanglement $E_N$(MC$_1$-YIG$_1$), $E_N$(MC$_1$-MC$_2$) and $E_N$(YIG$_1$-YIG$_2$) increase at first because the increase of the steady-state solution 
$\langle m_1 \rangle$ according to Eq(\ref{eq12}) leads to the increase of the effective coupling rate. After the initial increase $E_N$(MC$_1$-YIG$_1$), $E_N$(MC$_1$-MC$_2$) and $E_N$(YIG$_1$-YIG$_2$) decrease because $\langle A_1 \rangle$ decreases according to Eq(\ref{eq13}) and decreases the effective coupling rate. Meanwhile, the entanglement $E_N$(MC$_2$-YIG$_2$) increases because the entanglement is transferred to MC$_2$ as the entanglement in MC$_1$ decreases. The similar result is presented in Fig.\ref{fig8}(b) where we keep $\kappa_{m_2}$=0.1$\omega_b$ and increase $\kappa_{m_1}$.

\begin{figure}[!ht]
	\centering
	\includegraphics[width=0.49\linewidth]{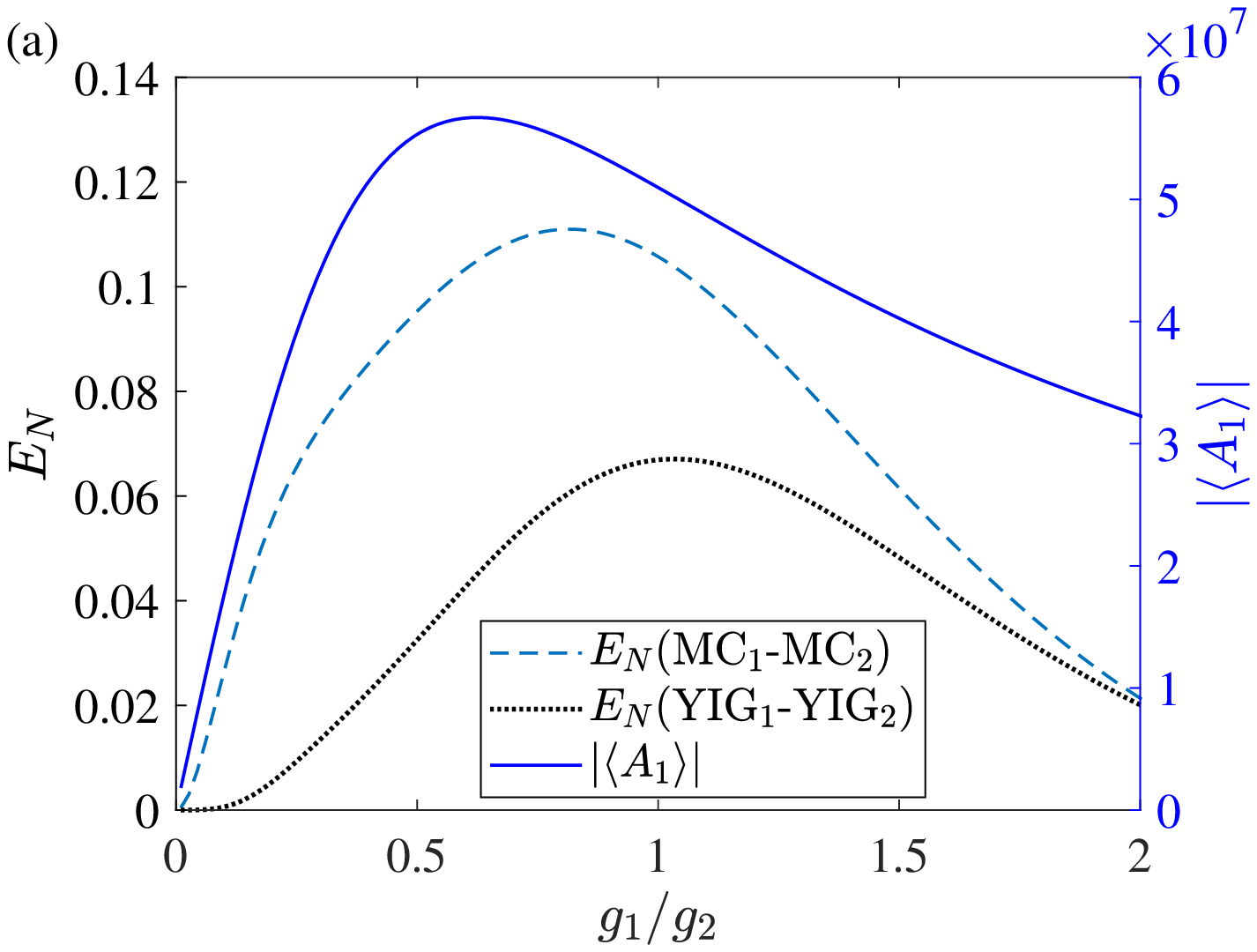}
	\includegraphics[width=0.49\linewidth]{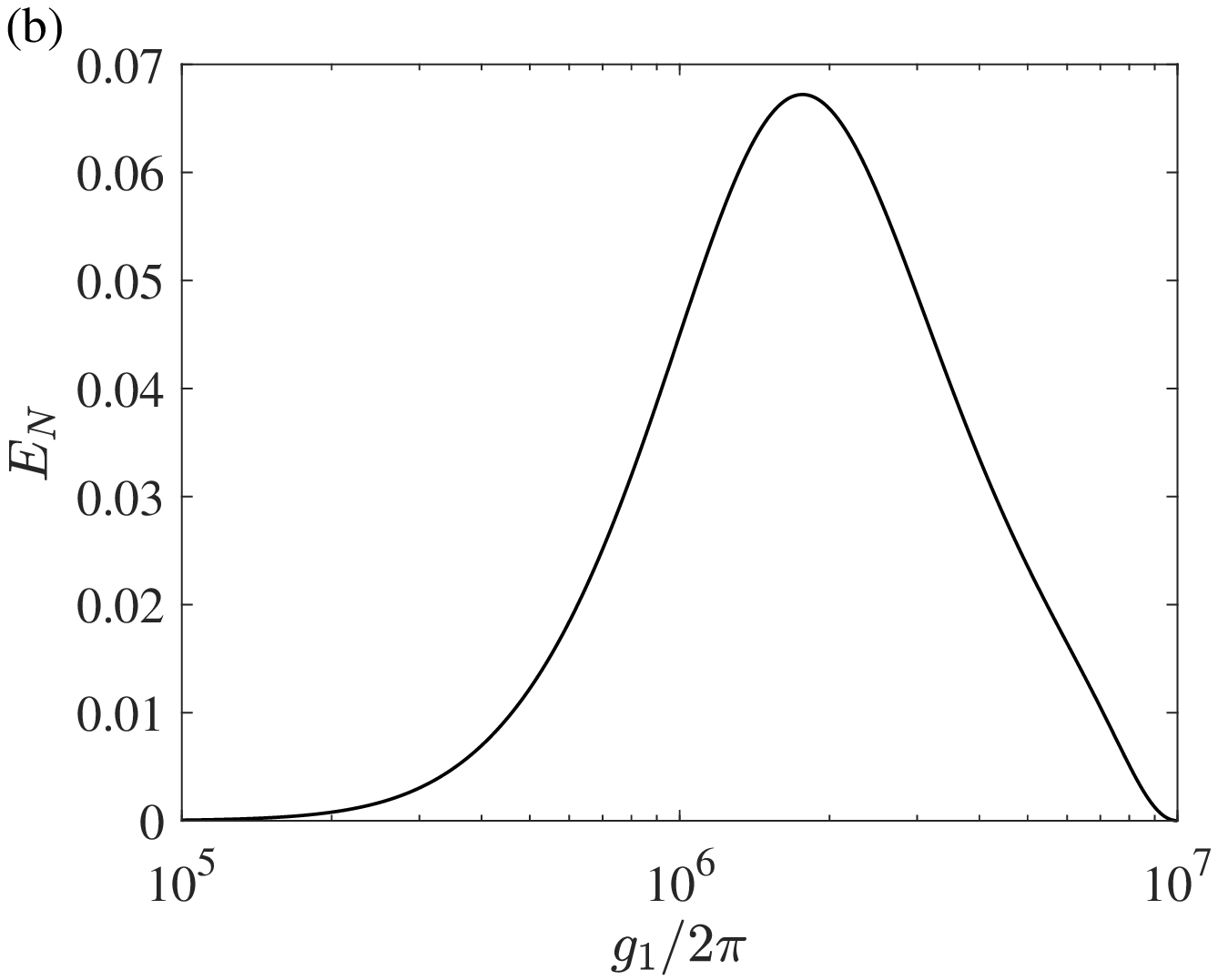}
	\caption{Entanglement $E_N$ versus the coupling rate between two YIG spheres. (a) $g_1$ varies individually while $g_2/2\pi$=1.7MHz. The variation of the steady state solution $\langle A_1 \rangle$ is shown with the right $y$ axis. (b)$E_N$(YIG$_1$-YIG$_2$) versus $g_1$ ($g_2$=$g_1$) which varies over a wide range. } \label{fig9} 
\end{figure}

The magnetic dipole coupling rate $g_1$ ($g_2$) between the MW cavity field and the YIG sphere is determined by the size and the position of the YIG sphere and can vary over a wide range \cite{Zhang2014}, therefore $g_1$ and $g_2$ are also important parameters to be optimized. Fig.\ref{fig9}(a) shows the variation of $E_N$(MC$_1$-MC$_2$) and $E_N$(YIG$_1$-YIG$_2$) versus $g_1$ while $g_2/2\pi$=1.7MHz. The entanglements increase at first and then begin to decrease. The variation of entanglement is determined by the effective coupling rate which is proportional to $\langle A_1 \rangle$. The variation of $\langle A_1 \rangle$ according to Eq(\ref{eq13}) is shown in Fig.\ref{fig9}(a). In addition, as the ultrastrong coupling has been realized\cite{Zhang2014}, we analyze the optimal value of $g_1$ over a wide range in Fig.\ref{fig9}(b) while $g_1$=$g_2$. The result shows that entanglement exists in a wide parameter regime and the optimal parameter is $g_1=g_2\approx1.7\times 2\pi$ MHz which matches the parameter that we have chosen in this work.

\section{Conclusion and Remarks}\label{sec4}

We present a novel cavity opto-magno-mechanical hybrid system to create entangled MW photons and YIG magnons, assisted by a cavity in the optical domain. Steady states are obtained by the red detuning of the optical driving field. Owing to the mechanical cooling process of the optical cavity, the system is robust against environmental temperature. We analyze the variation of entanglements versus different parameters and present the optimal condition. The two YIG spheres are embedded into two separate MW cavities respectively, therefore the ferromagnetic resonate frequencies of the two YIG spheres can be tuned independently, as well as the cavity frequencies. We analyze the individual tunability of the system and present the different entanglement features in the local-cavity and cross-cavity subsystems. The hybrid system provides a platform to generate entanglements between different physical systems. The entanglements of Gaussian states in this system represent two-mode squeezed states, which can be used to improve the precision beyond the standard quantum limit in quantum metrology \cite{zhang2014quantum, anisimov2010quantum}. In addition, the entangled Gaussian states can be used in quantum information tasks such as quantum teleportation \cite{wang2007quantum, braunstein2005quantum}. The individual tunability of the separated cavities allows us to independently control the entanglement properties of different subsystems and establish quantum channels with different entanglement properties in one system, and therefore provides potential applications in quantum information tasks.

\backmatter

\bmhead{Acknowledgments}

This work was supported by the National Natural Science Foundation of China (No. 12274037 and No. 61675028) and the  Interdiscipline Research Funds of Beijing Normal University. 

\bmhead{Data availability}

The datasets generated and analysed during the current study are available from the corresponding author on reasonable request

% \section{Section title of first appendix}\label{secA1}

%An appendix contains supplementary information.




%\end{appendices}

\bibliography{MyBib}


\end{document}